\documentclass[conference]{IEEEtran}
\usepackage{cite}
\usepackage{amsmath,amssymb,amsfonts}
\usepackage{algorithmic}
\usepackage{graphicx}
\usepackage{textcomp}
\usepackage{xcolor}
\usepackage{fancyhdr}
\usepackage[hyphens]{url}

\def\BibTeX{{\rm B\kern-.05em{\sc i\kern-.025em b}\kern-.08em
    T\kern-.1667em\lower.7ex\hbox{E}\kern-.125emX}}

\usepackage{booktabs}   
\usepackage{subcaption} 

\usepackage[belowskip=0pt,aboveskip=0pt]{caption}

\newcommand{\ie}{\textit{i}.\textit{e}.}
\usepackage[colorinlistoftodos,prependcaption,textsize=tiny]{todonotes}

\usepackage{tabularx}
\usepackage{array}

\usepackage{pgfplots,pgfplotstable}
\usepackage{tikz}
\usetikzlibrary{patterns}
\usepackage[section]{placeins}

\pgfplotsset{
        table/search path={figs},
    }

\definecolor{color_blue}{RGB}{37,103,185}
\definecolor{color_red}{RGB}{200,0,122}
\definecolor{color_green}{RGB}{80,101,10}

\definecolor{color_orange}{RGB}{0,155,0}
\definecolor{color_brown}{RGB}{160,83,1}
\definecolor{color_pink}{RGB}{18,46,208}
\definecolor{color_purple}{RGB}{158,38,217}
\definecolor{color_navy}{RGB}{190,10,10}

\definecolor{color_black}{RGB}{0, 0, 0}

\newcolumntype{C}[1]{>{\hsize=#1\hsize\centering\small\arraybackslash}X}%

\usetikzlibrary{external}
\title{NEAT: A Framework for Automated Exploration of Floating Point Approximations}
\author{\IEEEauthorblockN{Saeid Barati}
\IEEEauthorblockA{\textit{Computer Science Department} \\
\textit{University of Chicago}\\
Chicago, USA \\
saeid@cs.uchicago.edu}
\and
\IEEEauthorblockN{Lee Ehudin}
\IEEEauthorblockA{\textit{Computer Science Department} \\
\textit{University of Chicago}\\
Chicago, USA \\
ehudinl@uchicago.edu}
\and
\IEEEauthorblockN{Henry Hoffmann}
\IEEEauthorblockA{\textit{Computer Science Department} \\
\textit{University of Chicago}\\
Chicago, USA \\
hankhoffmann@cs.uchicago.edu}
}

\begin{document}
\maketitle
\thispagestyle{plain}
\pagestyle{plain}

\begin{abstract}
Much recent research is devoted to exploring tradeoffs between computational accuracy and energy efficiency
at different levels of the system stack.
Approximation at the floating point unit (FPU) allows saving energy by simply reducing the number of computed floating point bits in return for accuracy loss.
Although, finding the most energy efficient approximation for various applications with minimal effort is the main challenge.
To address this issue, we propose NEAT: a pin tool that helps users automatically explore the accuracy-energy tradeoff space induced by various floating point implementations.
NEAT helps programmers explore the effects of simultaneously using multiple floating point implementations to achieve the lowest energy consumption for an accuracy constraint or vice versa. NEAT accepts one or more user-defined floating point implementations and programmable placement rules for where/when to apply them. NEAT then automatically replaces floating point operations with different implementations based on the user-specified rules during the runtime and explores the resulting tradeoff space to find the best use of approximate floating point implementations for the precision tuning throughout the program.
We evaluate NEAT by enforcing combinations of 24/53 different floating point implementations with three sets of placement rules on a wide range of benchmarks. We find that heuristic precision tuning at the function level provides up to 22\% and 48\% energy savings at 1\% and 10\% accuracy loss comparing to applying a single implementation for the whole application. Also, NEAT is applicable to neural networks where it finds the optimal precision level for each layer considering an accuracy target for the model.

\end{abstract}



\maketitle

\section{Introduction}
Early work in approximate computing demonstrates the tremendous energy and
execution time reductions by making a variety of arithmetic and logic functional units available \cite{earlywork,CMOS,Palem4,DisciplinedProgramming}. Reduced-precision methods advocate less numerical precision for the data storage and computation to achieve higher performance and energy efficiency \cite{PrecisionScaling, wang2018, das2018,sakr2019}.

The proliferation of both different approximate functional units and reduced-precision software methods creates tremendous
opportunity, but it also creates a new problem.  While designing for reduced precision has long been common in specialized application domains---for example, digital signal processing \cite{Boutros2018}---the proliferation of these techniques means that general programmers will now have to consider the implications of such designs.
Specifically, it is up to programmers decide which level
of approximation to use at different points in their application and navigate through this immense tradeoff space enacted by allowing multiple approximations within a single program.

Consider 10 different levels of approximation available to be enforced at the function level for a moderate-sized  program with 10 functions. Programmers attempting to design for energy efficiency and accuracy in this scenario face two separate, but related, challenges.  First, is the challenge of correctly (in terms of achieved accuracy) implementing 10 different versions of each candidate function (one version for each available level of precision).  Second is the challenge of searching the resulting tradeoff space with ${10}^{10}$ points to explore. The tradeoff space could be even larger if we  exploit data type approximation where each variable in the program could acquire a different level of approximation\cite{fousse2007, Sampson2011, bornholt2014, tagliavini2018}.
Constructing a large number of alternative implementations and then navigating such an immense tradeoff space is likely beyond the abilities of even domain experts.  Thus, we need an automated precision tuning framework that can both generate alternative implementations and then explore the induced tradeoff space.

In this paper, we propose one mechanism that helps address both of the above challenges: \emph{programmable placement rules for approximate floating point computation}. We argue that asking programmers to implement $N$ different versions of key functions is unnecessarily burdensome and generating all possible approximations of each function will make the search space prohibitively large.  The programmable rules are a compromise, where programmers can encode their knowledge of the application into concise rules about which functions can be approximated, by how much, and when it might be permissible to do so.  These rules can then be used by an automated tool to generate a candidate set of approximate function implementations which is much smaller than the set of all possible approximations.

To address the challenges of creating and selecting from a large number of approximation alternatives, we propose NEAT---Navigating Energy Approximation
tradeoffs---a tool that helps users explore different levels of approximation within a program without detailed instrumentation and without laboriously creating many alternative implementations of functions.
NEAT accepts a user program, a set of approximate floating point implementations, and a set of
programmable placement rules for when to use a specific implementation within a program.
NEAT then runs the program and dynamically replaces floating point operations (FLOPs) with
the approximate version as specified by the rules.  NEAT reports the program's output with the estimation of floating point unit (FPU) and memory access energy alongside an itemized report of FLOPs in the program. Thus, NEAT helps developers explore the configuration space of
floating point implementations (FPI) without requiring them to have deep numerical expertise.

We implement NEAT for x86 using the Pin binary instrumentation system \cite{Luk05}.  We demonstrate
NEAT's value by comparing the approximations produced by different placement rule
sets.  In the first, we write a simple rule that  picks a single floating point implementation for the entire program; \ie{}
the rule is a simple one-to-one replacement (whole-program rule) common to many proposed approximation methods; e.g., those that use a single, reduced precision for machine learning \cite{Du2014} or scientific simulation \cite{Palem4}. In the second, we allow the top 10 executed functions with the most FLOPs to each use a different approximation (per-function rules). Either we use the currently-in-progress function (CIP) or the most recent function on the call stack (FCS) as the target to apply the approximate floating point implementation. For all rules, NEAT uses a genetic algorithm to guide exploration of the enormous resulting search space.

We evaluate NEAT on a selected set of benchmarks from Parsec 3.0\cite{Bienia2011} and Rodinia 3.1\cite{Che2009} suites which covers a variety of real-world applications. For the FPIs, we applied mantissa bitwidth tuning. On average, the per-function placement retrieves more energy-optimal floating point implementations than the whole-program approach, providing 22.1\% and 3.2\% energy savings in FPU and memory respectively with an allowance of 1\% accuracy loss. To ensure the robustness of NEAT, we include multiple inputs for each application which are divided into training
and test sets to evaluate whether NEAT produces statistically sound results. We also extend the evaluation by including a digit recognition application that is implemented with a neural network and MNIST dataset. For any accuracy target, NEAT provides the required precision level for each layer. NEAT is also released as opensource, so others could evaluate or use it freely.

In summary, this paper proposes:
\begin{itemize}
\item The NEAT framework that helps users explore the tradeoff space of reduced precision floating point combinations while not requiring hand tuning or code instrumentation.
\item A case study that compares whole-program vs. per-function approximation placements for a variety of benchmarks. Also, NEAT offers a separated placement solution based on the caller function, useful for the high frequency invoked functions.
\item Robustness on unseen inputs with a high correlation coefficient. NEAT finds statistically meaningful approximations that are not sensitive to input data and are more likely to be efficient on an unseen set of inputs.
\item A demonstration of NEAT's applicability to Convolutional Neural Networks (CNN), providing precision computation modes per layer resulting in energy savings with minimal loss of model accuracy.

\end{itemize}

\section{Background \& Motivation}

\subsection{Prior Work}
\label{related_work}

While there has been a substantial amount of effort aimed towards finding new forms of approximation \cite{Omer,Patterns2010,ICSE2010,Hoffmann2009,LoopPerforation,Hoffmann2011,Flikker,Memoization,Relax,Axilog,Parrot,Palem2,Sampson2011}, there is a lack of solutions that
helps the user to both develop their own approximation methods, and then specifying the approximation level to enforce for a single application.

Hardware approximation computes inexactly in return for
reduced energy, area, or time \cite{lingamneni2013,Chippa2013}.
Approximate multipliers \cite{Zervakis16,Khaing10,Kulkarni11} and adders \cite{Du12, Zhu10} are widely advocated for energy-efficient computing.
State-of-the-art neural network training platforms offer 16 bit floating point hardware systems that provide up to 4x performance gain comparing to traditional 32 bit systems \cite{Fleischer2018}.
Recent proposals promote putting many different approximate units or customized accelerators on a single core
\cite{MultipleFpu}. Thus, it is beneficial to include multiple FPUs on a chip for higher energy efficiency\cite{Palem4} but this requires tedious hand-tuning. Therefore, the challenge is how to figure out which FPU to use in each part of the program.  This is the challenge that motivates NEAT.

Languages support approximation allowing the specification of variants for key functionality and formal
analysis of their effects \cite{ansel2011,bornholt2014,chisel}. Approximation Knobs provide a way to lend performance and energy gains to existing
power knobs\cite{Knobs}. Quora is a quality programmable processor where the notion of quality is
codified in the instruction set of the processor \cite{PrecisionScaling}. Another example of user-defined approximation is Green, which is a system that allows programmers to supply approximate versions of loops and while-blocks that terminate early \cite{Baek2010}.
On the contrary to these programming language techniques, our proposal lets users easily---through our programmable substitution rules---examine and change the accuracy of FLOPs, giving them more control over the floating point computations in a program.

Performing precision tuning at fine grain is available through software libraries.  EnerJ proposes to declare approximate data via type qualifiers\cite{Sampson2011}.  MPFR adds to its arbitrary-precision representation the support for rounding modes, exceptions and special  values  as  defined  in the  IEEE  754  standard\cite{fousse2007}. FlexFloat reduces floating point emulation time by providing a C/C++ interface for supporting multiple FP formats. These techniques require source code instrumentation (changing $float$ and $double$ variables definition to custom parameters) or intending to yield more precise computation (for instance floating points numbers with more than 128 bits). NEAT focuses on energy efficiency by reducing precision while only requiring the program binary.

Convolutional Neural Networks (CNNs) include a significant amount of floating point computation in the training and inference stages. A large body of research has been focused towards CNN precision scaling\cite{sakr2019,Venkataramani2014,Zhang2015,Grigorian2015,Courbariaux2015,Han2016}. For example, WAGE quantizes weights to 2 bits while activation, errors, and gradient are 8 bits respectively\cite{wu2018}. FLEXpoint presents a new format with 16 bits mantissa to train CNNs with full precision\cite{Koster2017}. Another piece of research demonstrates the successful training with 8-16 bits floating point numbers with full accuracy\cite{wang2018}. Other, tangentially related approaches create networks with early exit points \cite{ALERT1,ALERT2}, but those are not related to the problem of changing numerical precision. Prior approaches either change the training architecture or apply a coarse-grain precision level for all layers. Differently, NEAT generates precision tuning analysis at different granularities by offering WP and CIP solutions without modifying the application internal structure or exhaustive precision exploration.

While prior work mainly develops \emph{mechanisms} that enable approximation to provide energy and runtime savings at different domains, they do not help users make more informed decisions about approximation. These techniques mostly are not flexible about how much, where, and when to approximate, and only provide discrete approximation knobs which leads to more conservative design choices. NEAT does not propose new mechanisms but helps users answer the questions above.

\subsection{Motivation}
\label{motivation}

\begin{figure}[t]
    \centering
    \includegraphics[width=1\linewidth]{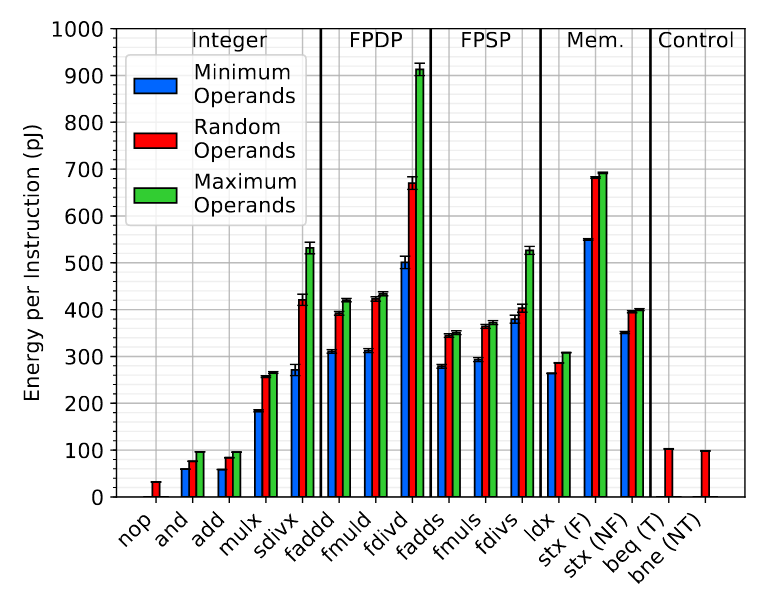}
    \caption{Energy Per Instruction for different classes of instructions.}
    \label{fig:epi_chart}
\end{figure}

Current inexact functional units in addition to approximate software libraries create an opportunity to  exploit  quality-energy  tradeoffs.
While an FPU accounts for 2-5\% area on the chip, the floating point instructions consume significantly more energy compared to other classes of instructions such as integer, memory, and control\cite{McKeown18, balkind2016}. Figure \ref{fig:epi_chart} illustrates the energy per instruction (EPI) results for different classes of instructions of 64-bit 32nm processor. With random operands, a 64-bit floating point add consumes 400 pJ, and a division operation could go as high as 680 pJ. For a 32-bit versions, the energy consumption is 350 and 420 pJ respectively.

As expected with regards to the type of operations, executing the floating point instructions emerges as a major contributor to the total energy consumption. Recent empirical studies have shown up to 50\% of the energy consumed in a core and memory is related to floating point instructions\cite{McKeown18}. Thus, exploiting reduced bitwidth at instruction level (bit truncation) to generate Floating Point Implementations (FPI) could facilitate higher energy efficiency. Another useful insight from Figure \ref{fig:epi_chart} is the relationship between computation and memory accesses. For example, three \texttt{add} operations consume the same amount of energy as a \texttt{ldx} instruction. Hence, looking from an energy efficiency point of view, reducing the memory traffic could be as efficient as optimizing the floating point arithmetic operations\cite{McKeown18}.

A body of literature has focused on providing tool supports that allow users to define several approximations for different components of the application \cite{EnerJ,Relax,Accept,Parrot,Hoffmann2011,Patterns2010,ICSE2010,Hoffmann2009}. Petabricks provides language  extensions that expose tradeoffs between time and accuracy to the compiler\cite{ansel2011}. The compiler then runs dynamic autotuning  to generate optimized elements to achieve the target accuracy. However, autotuners need to be determined on a per-application basis by the user. OpenTuner provides fully-customizable configuration representation and ensembles  of  search  techniques  to find an optimal solution\cite{Ansel2014}. Both autotuning techniques are supposed to help programmers but Petabricks requires a separate language and both require users to implement all alternatives before the search can be conducted. NEAT also helps users deal with approximation, but instead of requiring users to implement all possible alternatives, they simply describe programmable rules that are then used to automatically generate the alternatives.

Hence, there is a need for a generic framework that provides multiple precision levels, accommodates custom user-defined floating point implementation, and does not require code refactoring. NEAT provides such a solution. NEAT generates insightful information for precision tuning at function level for floating point programs.

\section{System Design}
\label{framework}


In this section, we describe our solution which generates insightful information about floating point precision tuning for applications. This tool, named Navigating Energy and Accuracy tradeoffs (referred to as NEAT) allows users to collect energy and performance data from applications using custom implementations of floating-point arithmetic.

The main challenge of precision tuning is constructing the right configuration of floating point precisions for the application. This configuration space might be extremely large to fairly small, ranging in complexity from using a different floating point implementation for each dynamic floating point instruction, using a different implementation for different function calls, or just picking a single floating point implementation for the entire application.
NEAT provides such flexibility in the granularity of enforcing floating point approximations by introducing the programmable placement rules and then automatically searching the accuracy and energy tradeoff space to find the optimal frontier.

\begin{figure}[t]
    \centering
    \includegraphics[width=\linewidth]{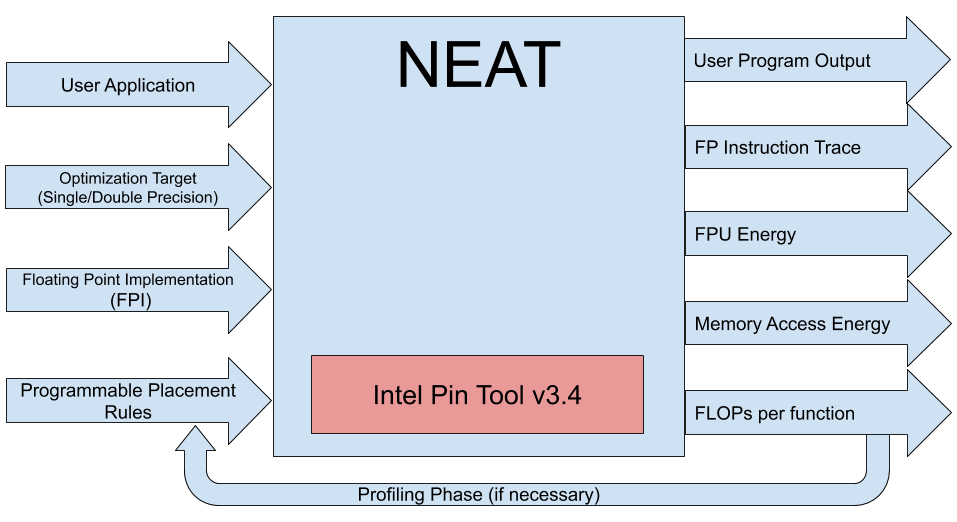}
    \caption{NEAT Design}
    \label{fig:design}
\end{figure}

Figure \ref{fig:design} illustrates the NEAT system from the user perspective. Users specify: (1) the application that they want to understand (this could be just a binary and requires no special changes), (2) whether NEAT should consider double or single precision (or both), a set of alternative implementations for floating-point arithmetic, and (4) the programmable placement rules that describe when, where, and how in the program to replace the standard floating point operations with one of the alternative implementations. NEAT then runs the program as a pin tool and intercepts all floating point operations of the specified type and replacing them according to the rules.  NEAT will perform multiple runs of the application, collect statistics on floating point usage, accuracy, and estimated energy. NEAT offers a profiling mode where the user collects precision analysis such as quantity and frequency of FLOPs for the application before applying any FPIs. Ultimately, NEAT can repeatedly test different assignments of floating point operations to find the frontier of optimal configurations; i.e., assignments of floating point operations to different regions of the code.
This section describes NEAT's inputs, internals, and outputs.


\subsection{NEAT Inputs}
\label{NEAT_inputs}

User inputs of NEAT includes: a user application to instrument, a precision level as the optimization target, the desired FP arithmetic implementations, and a set of FPI to function mappings (programmable placement rules).

NEAT receives the binary of the program and instruments the floating point instructions. Unlike other precision tuning tools, NEAT does not require the source code of the program. Then, NEAT expects the optimization target which can be either single or double precision. There are two reasons behind including optimization objective. First, for most of the programs, the same precision level is held across the code base for the data structures and the functions. Second, if we consider both $float$ and $double$ FLOPs to optimize, the configuration space of FPIs combinations would explode excessively.

Next, users specify multiple FPIs for any individual arithmetic instruction such as addition, subtraction, multiplication, and division for each operand. At last, NEAT expects a mapping between the candidate code sections and the FPIs to calculate each FLOP in a program. By default, NEAT enforces the FPIs at the function level, meaning all FLOPs executed within a specific function will be using the same customized FPI. Any function that has at least one FLOP can be considered as a candidate for approximation.

\subsection{NEAT Internal Structure}
\label{internal_structure}

The NEAT dynamic instrumentation tool was written in
C++ using the Intel Pin instrumentation system \cite{Luk05}. NEAT performs run-time instrumentation to facilitate the analysis and replacement of floating-point arithmetic operations during the execution of compiled C and C++ binaries.

\subsubsection{Intel Pin Tool}
\label{pintool}

The Pin instrumentation system was chosen as the backbone for this tool because of its clean API and efficient implementation. The Pin API makes it possible to write instrumentation routines to observe and alter the architectural state of a process. Pin uses a JIT compiler to generate a new instrumented code that can be executed without the extra runtime overhead from instrumentation.

\subsubsection{Floating Point Operations}
\label{flops}

For the purposes of this tool, we identify floating-point
arithmetic operations as the Streaming SIMD Extensions
(SSE) instructions for scalar arithmetic. These instructions
are included in a SIMD instruction set extension to the x86
architecture and operate on 32-bit or 64-bit floating
point numbers. More specifically, the instructions we use for
our definition of floating-point operation are ADDSS, SUBSS, MULSS, DIVSS, ADDSD, SUBSD, MULSD, and DIVSD.

\subsubsection{Floating Point Arithmetic Implementation}
\label{fpi}

Custom hardware units or accelerators have been considered for enriching the quality versus energy tradeoff spaces. Approximate adders \cite{Verma08,Zhu10,Du12} and multipliers \cite{Kulkarni11, Khaing10,Zervakis16} have been designed as a solution for lower power consumption and high performance. In the presence of inexact hardware units, NEAT provides information on how to efficiently redirect the arithmetic instructions to these units.

The floating point formats with a lower number of bits emerge an appealing opportunity to reduce the energy consumption since it allows simplification of both hardware units and reduction of memory bandwidth required to transfer the data between the memory and registers. The FPI can be as simple as bit truncation in the FP format representation, enforcing direct approximation to the operands or result of arithmetic operations, or redirecting instruction to approximate hardware units or software libraries.

\subsubsection{Execution of Floating Point Instructions}
\label{exec_flops}

Defining an FPI is fairly trivial. The main challenge with enforcing FPI dynamically is the way to specify the exact mapping between FPI and the FLOPs. NEAT allows users to define placement rules that determine which FPI is used to calculate each FLOP in a program. Every time a FLOP is about to be calculated in the user application, NEAT examines all of the mappings and captures information about the current state of the application, and use them to determine
which FPI will be applied to calculate the result of the FLOP.

\def\tabularxcolumn#1{m{#1}}
\newcolumntype{D}[1]{>{\hsize=#1\hsize\centering\bfseries\footnotesize\arraybackslash}X}%
\newcolumntype{E}[1]{>{\hsize=#1\hsize\centering\footnotesize\arraybackslash}X}%
\newcolumntype{F}[1]{>{\hsize=#1\hsize\centering\tiny\arraybackslash}X}%

\begin{table}[tb]
  \centering
  \caption{Built-in Placement Rules in NEAT.}
\footnotesize
\begin{tabularx}{0.45\textwidth}{E{0.7}|E{1.6}|E{0.7}}
 \textbf{Placement Rule} & \textbf{Description} & \textbf{tradeoff Space Size} \\
\hline
WP & one FPI for the whole program                          & $24 - 53 $\\
CIP &  one FPI for the currently in progress function           & $24^{10} - 53^{10}$ \\
FCS & one FPI for the most recent function on the call stack & $24^{10} - 53^{10}$   \\
\end{tabularx}
\label{Tab:neat_placement_rules}
\end{table}

NEAT comes packaged with three predefined sets of FPI placements for the applications the cover many use-cases and show off its versatility. Table \ref{Tab:neat_placement_rules} includes the default placement rules and the corresponding tradeoff space size. Sets of rules are specified as C++ routines that accept the program state as input and return a single FPI as output.

The first set applies the same FPI for every FLOP in the whole-program (WP) regardless of the current function and the program state.

For finer granularity, the user can register callbacks through NEAT that can be executed whenever a function is entered or exited in the instrumented application. This allows more complex information to be collected about the program state, such as the call stack of the application.
The second set of placement rules allows the user to specify a map of function names to FPIs and employs each FPI for the FLOPs in the corresponding currently in progress (CIP) function.
Similarly, the third set of placement rules uses callbacks registered with NEAT to keep track of the function call stack (FCS) of the program. Instead of inspecting the current function, NEAT first checks the most recent function on the call stack.  If no functions in the call stack match the names of those in the user-supplied map, a default implementation is used.

\begin{figure}[t]
    \centering
    \includegraphics[width=0.75\linewidth]{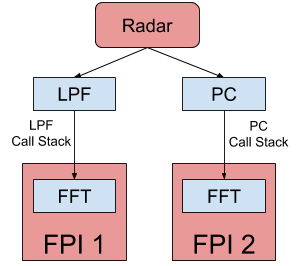}
    \caption{FCS placement considers FFT function call stack before selecting the approximate FPI.}
    \label{fig:radar_design}
\end{figure}

To highlight the difference between CIP and FCS, we analyzed the structure of 7 functions in a benchmark shown in Figure \ref{fig:radar_design}. The \texttt{radar} is an embedded real-time signal processing application that is used to find moving targets on the ground \cite{Lebak05,Hoffmann2012}. It includes both a low-pass filter (LPF) and pulse compression (PC). Both of these components use a Fast Fourier Transform (FFT) as a part of their computation.

With the CIP option, NEAT enforces the same FPI every time the FFT function is called. For the FCS option, NEAT distinguishes between the two occurrences of FFT based on who has made the function call. Therefore, NEAT uses one FPI for the FFT in the Low Pass Filter (LPF) stage and a second FPI for the FFT in the Process Pulse (PC) stage.
Empirically, we have found the results of FCS and CIP for most of the benchmarks do not differ as the callers of a FLOP intensive functions are the same. The \texttt{radar} is an example where multiple functions make numerous calls to the same FLOP-intensive function that is accuracy sensitive.

\subsection{Outputs}
\label{NEAT_outputs}

There are five outputs from this tool: the output from the
user application, a trace of the operands and result of every
FLOP executed by the program, the estimated FPU energy of FLOPs in the execution of the program, the estimated energy of off-chip memory accesses of the program, and the number of
FLOPs executed per function in the program.

The trace of the FLOPs executed by the instrumented application is written to a file while the application is running. If FPIs are supplied to NEAT by
the user, the result of each operation will be printed after the
operation is calculated with the chosen FPI. The operands and result of each operation are printed as hexadecimal numbers so that there is no confusion
in rounding the floating-point values.

NEAT reports total energy consumed in FPU by using energy per instruction (EPI) of different classes floating-point operations. We extracted the energy model of $fadd$, $fmul$ and $fdiv$ for single and double precision operations provided in related work \cite{McKeown18}.

To this end, NEAT counts the number of bits manipulated in the operands and results of every FLOP in the instrumented program. Modifying the bit width in the exponent and sign of a floating-point number changes the accuracy significantly where the quality of output becomes unacceptable. Hence, NEAT only focuses on mantissa bits. NEAT counts the number of zeroes in the binary representation of the floating-point number, starting with the least significant bit, and then subtracts it from available mantissa bits in the floating type (24/53 bits in single/double precision respectively) to calculate the number of manipulated bits. NEAT uses the EPI models and the number of manipulated bits per FLOP to estimate the total floating-point energy consumed in the FPU.

NEAT also records the total number of bits used in FLOPs in the execution
of the program is output to a file after the termination of the application. Unlike the FPU energy estimation, this metric can be used
as a platform-independent way to evaluate the approximate
amount of power used by FLOPs when instrumenting a program.

Currently, the memory accounts for more than 25\% of energy spent in a large scale system. While on average, each single precision FLOP takes 400 pJ to execute, a byte read from memory consumes 1.5 nJ \cite{Borkar11}. Accordingly, NEAT counts the total number of bits transmitted to/from memory and then estimates the total memory access energy of the instrumented program \cite{Malladi12}. This allows NEAT to yield a better energy estimation of the program in a real system.

NEAT generates in-detail statistics about the floating point instructions in the program. Users might operate NEAT to profile the application before performing precision tuning to first, decide whether NEAT is useful to their application and second, what type of FPIs, which functions, and how to map them.

In general, NEAT is a tool used at program design time. NEAT allows users to evaluate many points on the accuracy/energy tradeoff curve without having to implement all possible alternatives.  After profiling with NEAT, users can then select a point and implement it with confidence that it will provide the desired behavior.

Future work would explore additional machione learning techniques to
configure the floating point usage differently for different functions
in the program \cite{ding2019,ESP,LEO,Imes2018}.  Another promising
line of work is using a runtime system to dynamically tune floating
point usage to maintain either energy or accracy constraints in a
changing workload
\cite{Copper,caloree,POET,FSE2015,FSE2017,coadapt,Meantime,TAAS2017,ASPLOS2018,bard,proteus},
or possibly implementing this control scheme in hardware
\cite{Grape,CASH,MERLOT}.
\section{NEAT Interface and Runtime}
\label{user_procedure}

We explain how the user can manage floating point precision scaling with the NEAT framework explained in the previous section. We specify the information that NEAT expects to receive from the users and then, discuss steps to execute the runtime engine of NEAT.

The NEAT procedure follows as:

1. \textbf{Profile the Program}: User runs the application. NEAT records the single and double precision instructions and the functions associated with them, and generates the detailed report in csv format.

2. \textbf{Assign FP Optimization Target}: Since the applications usually use the same precision level across the source code, NEAT enhances either single or double precision instructions at the same time. At this point, the user defines the directive for NEAT to target 32 or 64 bit FLOPs.

3. \textbf{Develop FPIs}: Users might define multiple FPIs to be explored by NEAT. NEAT supports FPIs developed in a number of different ways.  An FPI can be created by truncating mantissa bits of the FLOP representation or injecting direct approximation to the operands or results of floating point arithmetic operations. For example, approximating the inverse function \cite{Zhang2014} or $sin$ function using a neural network\cite{Eldridge2014} is considered an FPI, too. The FPI can be applied to one or more floating point arithmetic instruction. For instance, one benchmark might include numerous accumulations but few divisions. Thus, the user defines an FPI with enforcing 8 precision bits for the add/sub arithmetic instructions and 24 precision bits for the multiply instructions. The user develops an FPI by creating an instance of  the $FpImplementation$ virtual class. Furthermore, user might customize the subroutine of $PerformOperation$ to modify the operands or results of a floating point instruction directly.

4. \textbf{Register FPI Placement Rules and Functions}: NEAT expects to receive a mapping between FPIs and when to enforce them. For the WP approach, the user only needs to instantiate $Register\_FP\_selector$ class with the desired FPI as the argument. For the per-function rules, NEAT by default considers the top 10 FLOP intensive functions. The user might pre-profile the program to detect and select any number of functions. The user then should provide a mapping between functions and FPIs by defining a $pair <functionName , FPI* >$ map data structure. Next, the user should combine the map with one of the pre-packaged placement rules (CIP or FCS).  This mapping is also referred as a \textit{configuration}. Finally, the user creates an instance of $Register\_FP\_selector$ class and passes the map and placement strategy as the input arguments. At the runtime, the user passes the registered instance name via \texttt{fp\_selector\_name} command line flag to NEAT. This interface is simple, but provides a quite flexible approach to replacing standard floating point operations with the approximate version.  For example, the user can provide several maps and then their instantiation of the selector class can look at the current program context to select the desired map.  This allows NEAT to explore many different options for a single function within a program.  For example, users can specify that the map should depend on the function call stack so that different FP implementations will be used for the same function based on where it was called from.

5. \textbf{Activate Exploration Scripts}: If CIP or FCS schema is selected, the tradeoff space of FPI to function mappings (configurations) becomes too huge to explore exhaustively. Hence, NEAT uses the NSGA-II genetic exploration technique to search for energy efficient configurations \cite{NSGA-II}. If the user desires to enhance the exploration phase of the configuration space further, NEAT provides an interface through the command line flags to manually modify the tuning parameters of NSGA-II such as population size, number of generations, or convergence threshold.

6. \textbf{Analyze the Output}: NEAT reports detailed energy and performance data per configuration. Moreover, a python script is provided to generate scatter plots of tradeoff space with the lower convex hulls.

At the completion of these steps, the user finds information about the most appropriate precision level for each individual function or the whole program.

\section{Experimental Results}
\label{evaluation}
We evaluate the efficacy and flexibility of NEAT to provide floating point approximation analysis. In general, NEAT generates useful information on precision tuning of applications which can be used at design stage of a software or convoyed to other layers of system such as compilers or hardware (\textit{e.g.} building a set of reduced-precision FPUs). Section \ref{ss:fp_PD} inspects the floating point profiling of NEAT for the applications.
The primary challenge of automatic precision tuning is creating approximation configurations. We examine the NEAT's flexibility to produce customized FPI definitions in Sections \ref{ss:fpu_energy_saving} and \ref{ss:mem_insns}. Moreover, the main mechanism of NEAT---programmable placement rules---are investigated in Sections \ref{ss:flex_precision_level} and \ref{ss:func_call_stack}.

To navigate through the immense configuration space, NEAT comes with a tunable genetic exploration algorithm which is used in Sections above (from \ref{ss:fp_PD} through \ref{ss:fpu_energy_saving}). Although, to ensure robustness of NEAT on unseen data, we evaluate the difference between predicted accuracy and energy on training and test data to demonstrate that NEAT finds configurations that are robust across different test inputs that were not seen in training \ref{ss:coeff_cov}. Finally in section \ref{ss:nn_integration_eval}, we evaluate NEAT's general applicability to find appropriate reduced precision floating point configurations by evaluating it on a problem that has seen a tremendous amount of attention from human experts recently: trading accuracy for precision in neural network inference.  We find that NEAT
can use the whole-program rule to automatically find a single floating point precision that is similar to those reported by human experts.  Further, we find that by using different floating point implementations for different layers, NEAT produces even greater energy savings for the same accuracy.




\subsection{Evaluation platform}
\label{ss:eval_platform}

We evaluate NEAT by exploring the tradeoff spaces of the placement rules for a variety of benchmarks. Table \ref{tbl:benchmarks} lists the applications from Parsec 3.0 \cite{Bienia2011} and Rodinia 3.1 \cite{Che2009} suites with the configuration space size (default precision optimization target) and training and test inputs for each benchmark. These benchmarks cover domains from finance to image processing.

\newcolumntype{E}[1]{>{\hsize=#1\hsize\centering\tiny\arraybackslash}X}%
\newcolumntype{M}[1]{>{\centering\arraybackslash}m{#1}}

\begin{table*}[tb]
  \centering
  \caption{Benchmarks Used for Evaluation.}
\footnotesize
\begin{tabularx}{0.95\textwidth}{|C{0.8}|C{1.1}|C{1.1}|C{1}|}
\hline
 \textbf{Benchmarks } & \textbf{Training inputs} & \textbf{Test inputs} & \textbf{Possible Configuration Space}\\
\hline
Blackscholes  & 10 lists with 100K initial prices & 30 lists with 100K initial prices & $24^4$ \\
\hline
Bodytrack  & Sequence of 5 frames & Sequence of 20 frames & $24^{24}$ \\
\hline
Fluidanimate  & 5 fluids with 15K+ particle & 15 fluids with 15K+ particle & $24^9$ \\
\hline
Ferret  & 5 databases of 16 images & 15 databases of 16 images & $24^{12}$ \\
\hline
Heartwall & Sequence of 15 frames & Sequence of 60 frames & $24^4$ \\
\hline
Kmeans & 10 vectors with 512 data points & 30 vectors with 512 data points & $24^9$ \\
\hline
Particlefilter & Sequence of 32 frames & Sequence of 128 frames & $53^{10}$ \\
\hline
Radar & Sequence of 10 frames & Sequence of 40 frames & $24^{13}$ \\
\hline
\end{tabularx}
\label{tbl:benchmarks}
\end{table*}

To create FPIs, we use bit truncation. For the single precision floating point numbers ($float$ type in C), we have 24 different FPIs corresponding to the mantissa bits. Similarly, we created 53 FPIs for the double precision floating point numbers. For the whole-program approach, the size of the tradeoff space is the total number of possible FPIs which are 24 and 53 points. For the per-function approaches, we consider the top 10 functions with most FLOPs to enforce the FP rules, so each of the top 10 functions may use a different FPI.

In each experiment, at most 400 configurations in the tradeoff space (less than $6^{-12}$ of all possible configurations) have been evaluated through NEAT's genetic algorithm.

\subsection{Floating Point Precision Distribution}
\label{ss:fp_PD}
\begin{figure}[tb]
  \begin{center}
\begin{tikzpicture}

    \pgfplotsset{
        height=4cm,
        width=0.95\columnwidth,
        enlargelimits=false,
    }

    \pgfplotsset{every axis legend/.append style={
    draw=none,legend columns=2,at={(0.5,1.3)},anchor=north, column sep=1ex, font=\tiny},
    }


   \begin{axis}[
        axis x line* = bottom,
        major x tick style = transparent,
        xmin=0,
        xmax=12,
        xticklabel shift={0pt},
        x tick label style={rotate=35, anchor=east},
        xtick={1,2,3,4,5,6,7,8,9,10,11},
        xticklabels={{\scriptsize $\mathsf{Blackscholes}$},
        {\scriptsize $\mathsf{Bodytrack}$},
        {\scriptsize $\mathsf{Canneal}$},
        {\scriptsize $\mathsf{Fluidanimate}$},
        {\scriptsize $\mathsf{Ferret}$},
        {\scriptsize $\mathsf{Heartwall}$},
        {\scriptsize $\mathsf{Kmeans}$},
        {\scriptsize $\mathsf{Particlefilter}$},
        {\scriptsize $\mathsf{Radar}$},
        {\scriptsize $\mathsf{Srad}$},
        {\scriptsize $\mathsf{Swaptions}$}},
        ybar=\pgflinewidth,
        bar width=5,
        ymajorgrids,
        grid style={dashed},
        ylabel={\small Ratio to all FLOPs },
        ymin=0,
        ymax=1,
        ytick={0,0.25,0.50,0.75,1},
        yticklabels={0,0.25,0.50,0.75,1},
      ]
        \addplot[color=color_orange,fill=color_orange] table[x index=0,y index=2, col sep=tab] {DataflopsBreakdown.txt};
        \addplot[color=color_pink,fill=color_pink] table[x index=0,y index=3, col sep=tab] {DataflopsBreakdown.txt};
        \addlegendentry{{\footnotesize $Single Precision (32 bits)$}}
        \addlegendentry{{\footnotesize $Double Precision (64 bits)$}}





    \end{axis}


\end{tikzpicture}
    \caption{Floating Point Type Breakdown for Benchmarks. While most benchmarks have a dominant FP type, some carry both.}
    \label{fig:fpDist}
  \end{center}
\end{figure}
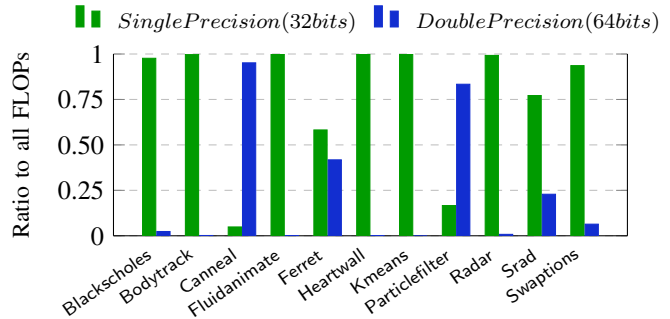

NEAT can be used to analyze the type, distribution, and the intensity of the FLOPs in a program. Figure \ref{fig:fpDist} depicts the ratio of single and double precision FLOPs for each benchmark.

Most of the benchmarks hold the same precision level across the source for correctness and portability. For example, \texttt{Bodytrack}, \texttt{Heartwall}, and \texttt{Kmeans} are all implemented with $float$ type while \texttt{Canneal} is mainly using $double$. However, for some benchmarks such as \texttt{Ferret}, \texttt{Particlefilter}, and \texttt{Srad} due to including external libraries, there is a mixture of both precision levels. In this case, users might choose the optimization target to be enforced. Specifying the right target opens up further opportunities for additional energy savings.

\subsection{FPU Energy Saving}
\label{ss:fpu_energy_saving}


NEAT provides the FPU energy estimation consumed by the FLOPs. We compare two rules: whole program (WP) and currently-in-progress (CIP).  As a reminder, WP uses one floating point implementation through the entirety of the program, while CIP is free to choose a separate implementation for each of the top 10 functions (by FLOP count) in the program.  For \texttt{Particlefilter}, we set the optimization target to double precision  as most of the FLOPs are $double$. For the rest of the benchmarks, we apply the single precision optimization.

We consider the top 10 FLOP intensive functions for the CIP placement. Although, one might ask how much of the FLOPs are included in the top 10 functions. For all benchmarks, at least 98\% FLOPs were coming from the top 10 functions, thus NEAT covers almost all of the FLOPs in the program.

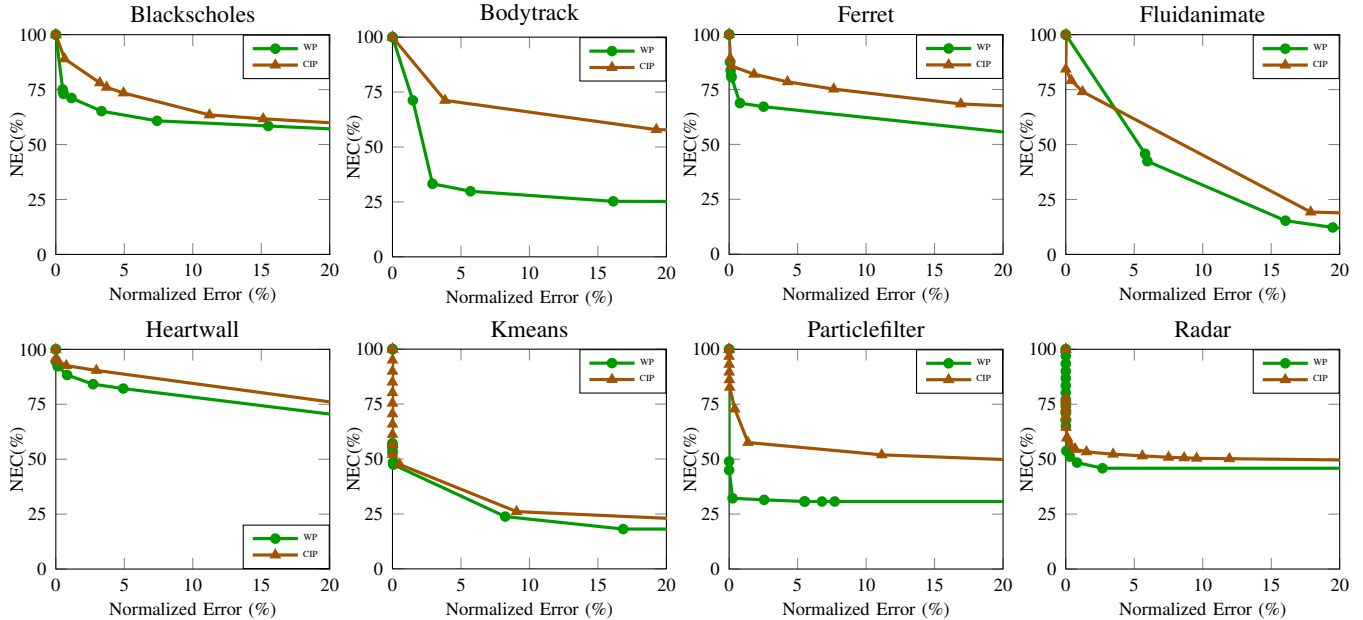
\begin{figure*}[t]
     \centering
     \begin{subfigure}{.24\linewidth}
         \tikzset{mark options={mark size=1.4 pt}}
%
\pgfplotsset {every mark/.append style={line width=0.4pt}}
\pgfplotsset{every axis/.append style={font=\scriptsize}}

\newcommand{\axisheight}{0.54408}


\begin{tikzpicture}

    \begin{axis}[
        title={\small {Blackscholes}},
        title style={yshift= -0.5em},
        xlabel={Normalized Error (\%)},
        ylabel={NEC(\%)},
        xlabel style={yshift= 0.6em},
        y label style={at={(axis description cs:0.2,.5)}},
        xmin=0, xmax=20,
        ymin=0, ymax=100,
        xtick={0,5,10,15,20},
        xticklabels={0,5,10,15,20},
        ytick={0,25,50,75,100},
        yticklabels={0,25,50,75,100},
        legend pos=north east,
        width=1.2\linewidth,
        legend style={font=\tiny, nodes={scale=0.7, transform shape}, at={(1.00,1.00)},anchor=north east},
        grid style=dashed
    ]

    %

    \addplot[color=color_orange, line width=1.2pt, mark=*] table[x index=0,y index=1,col sep=comma] {src/data/LowerConvexHull_FPUEnergy/Blackscholes-CIP.csv};
    \addplot[color=color_brown, line width=1.2pt, mark=triangle*] table[x index=0,y index=1,col sep=comma] {src/data/LowerConvexHull_FPUEnergy/Blackscholes-WP.csv};

    \legend{WP, CIP}


    \end{axis}

\end{tikzpicture}

\vspace{-0.1em}
         \label{fig:fpu_energy_Radar}
     \end{subfigure}
     \begin{subfigure}{.24\linewidth}
       \tikzset{mark options={mark size=1.4 pt}}
%
\pgfplotsset {every mark/.append style={line width=0.4pt}}
\pgfplotsset{every axis/.append style={font=\scriptsize}}

\newcommand{\axisheight}{0.54408}


\begin{tikzpicture}

    \begin{axis}[
        title={\small {Bodytrack}},
        title style={yshift= -0.5em},
        xlabel={Normalized Error (\%)},
        ylabel={NEC(\%)},
        xlabel style={yshift= 0.6em},
        y label style={at={(axis description cs:0.2,.5)}},
        xmin=0, xmax=20,
        ymin=0, ymax=100,
        xtick={0,5,10,15,20},
        xticklabels={0,5,10,15,20},
        ytick={0,25,50,75,100},
        yticklabels={0,25,50,75,100},
        legend pos=north east,
        width=1.2\linewidth,
        legend style={font=\tiny, nodes={scale=0.7, transform shape}, at={(1.00,1.00)},anchor=north east},
        grid style=dashed
    ]

    %

    \addplot[color=color_orange, line width=1.2pt, mark=*] table[x index=0,y index=1,col sep=comma] {src/data/LowerConvexHull_FPUEnergy/Bodytrack-CIP.csv};
    \addplot[color=color_brown, line width=1.2pt, mark=triangle*] table[x index=0,y index=1,col sep=comma] {src/data/LowerConvexHull_FPUEnergy/Bodytrack-WP.csv};

    \legend{WP, CIP}


    \end{axis}

\end{tikzpicture}

\vspace{-0.1em}
         \label{fig:fpu_energy_Bodytrack}
     \end{subfigure}
     \begin{subfigure}{.24\linewidth}
       \tikzset{mark options={mark size=1.4 pt}}
%
\pgfplotsset {every mark/.append style={line width=0.4pt}}
\pgfplotsset{every axis/.append style={font=\scriptsize}}

\newcommand{\axisheight}{0.54408}


\begin{tikzpicture}

    \begin{axis}[
        title={\small {Ferret}},
        title style={yshift= -0.5em},
        xlabel={Normalized Error (\%)},
        ylabel={NEC(\%)},
        xlabel style={yshift= 0.6em},
        y label style={at={(axis description cs:0.2,.5)}},
        xmin=0, xmax=20,
        ymin=0, ymax=100,
        xtick={0,5,10,15,20},
        xticklabels={0,5,10,15,20},
        ytick={0,25,50,75,100},
        yticklabels={0,25,50,75,100},
        legend pos=north east,
        width=1.2\linewidth,
        legend style={font=\tiny, nodes={scale=0.7, transform shape}, at={(1.00,1.00)},anchor=north east},
        grid style=dashed
    ]

    %

    \addplot[color=color_orange, line width=1.2pt, mark=*] table[x index=0,y index=1,col sep=comma] {src/data/LowerConvexHull_FPUEnergy/Ferret-CIP.csv};
    \addplot[color=color_brown, line width=1.2pt, mark=triangle*] table[x index=0,y index=1,col sep=comma] {src/data/LowerConvexHull_FPUEnergy/Ferret-WP.csv};

    \legend{WP, CIP}


    \end{axis}

\end{tikzpicture}

\vspace{-0.1em}
         \label{fig:fpu_energy_Ferret}
     \end{subfigure}
      \begin{subfigure}{.24\linewidth}
        \tikzset{mark options={mark size=1.4 pt}}
%
\pgfplotsset {every mark/.append style={line width=0.4pt}}
\pgfplotsset{every axis/.append style={font=\scriptsize}}

\newcommand{\axisheight}{0.54408}


\begin{tikzpicture}

    \begin{axis}[
        title={\small {Fluidanimate}},
        title style={yshift= -0.5em},
        xlabel={Normalized Error (\%)},
        ylabel={NEC(\%)},
        xlabel style={yshift= 0.6em},
        y label style={at={(axis description cs:0.2,.5)}},
        xmin=0, xmax=20,
        ymin=0, ymax=100,
        xtick={0,5,10,15,20},
        xticklabels={0,5,10,15,20},
        ytick={0,25,50,75,100},
        yticklabels={0,25,50,75,100},
        legend pos=north east,
        width=1.2\linewidth,
        legend style={font=\tiny, nodes={scale=0.7, transform shape}, at={(1.00,1.00)},anchor=north east},
        grid style=dashed
    ]

    %

    \addplot[color=color_orange, line width=1.2pt, mark=*] table[x index=0,y index=1,col sep=comma] {src/data/LowerConvexHull_FPUEnergy/Fluidanimate-CIP.csv};
    \addplot[color=color_brown, line width=1.2pt, mark=triangle*] table[x index=0,y index=1,col sep=comma] {src/data/LowerConvexHull_FPUEnergy/Fluidanimate-WP.csv};

    \legend{WP, CIP}


    \end{axis}

\end{tikzpicture}

\vspace{-0.1em}
         \label{fig:fpu_energy_Fluidanimate}
     \end{subfigure}
     \begin{subfigure}{.24\linewidth}
         \tikzset{mark options={mark size=1.4 pt}}
%
\pgfplotsset {every mark/.append style={line width=0.4pt}}
\pgfplotsset{every axis/.append style={font=\scriptsize}}

\newcommand{\axisheight}{0.54408}


\begin{tikzpicture}

    \begin{axis}[
        title={\small {Heartwall}},
        title style={yshift= -0.5em},
        xlabel={Normalized Error (\%)},
        ylabel={NEC(\%)},
        xlabel style={yshift= 0.6em},
        y label style={at={(axis description cs:0.2,.5)}},
        xmin=0, xmax=20,
        ymin=0, ymax=100,
        xtick={0,5,10,15,20},
        xticklabels={0,5,10,15,20},
        ytick={0,25,50,75,100},
        yticklabels={0,25,50,75,100},
        legend pos=north east,
        width=1.2\linewidth,
        legend style={font=\tiny, nodes={scale=0.7, transform shape}, at={(1.00,0.00)},anchor=south east},
        grid style=dashed
    ]

    %

    \addplot[color=color_orange, line width=1.2pt, mark=*] table[x index=0,y index=1,col sep=comma] {src/data/LowerConvexHull_FPUEnergy/Heartwall-CIP.csv};
    \addplot[color=color_brown, line width=1.2pt, mark=triangle*] table[x index=0,y index=1,col sep=comma] {src/data/LowerConvexHull_FPUEnergy/Heartwall-WP.csv};

    \legend{WP, CIP}


    \end{axis}

\end{tikzpicture}

\vspace{-0.1em}
         \label{fig:fpu_energy_Heartwall}
     \end{subfigure}
     \begin{subfigure}{.24\linewidth}
       \tikzset{mark options={mark size=1.4 pt}}
%
\pgfplotsset {every mark/.append style={line width=0.4pt}}
\pgfplotsset{every axis/.append style={font=\scriptsize}}

\newcommand{\axisheight}{0.54408}


\begin{tikzpicture}

    \begin{axis}[
        title={\small {Kmeans}},
        title style={yshift= -0.5em},
        xlabel={Normalized Error (\%)},
        ylabel={NEC(\%)},
        xlabel style={yshift= 0.6em},
        y label style={at={(axis description cs:0.2,.5)}},
        xmin=0, xmax=20,
        ymin=0, ymax=100,
        xtick={0,5,10,15,20},
        xticklabels={0,5,10,15,20},
        ytick={0,25,50,75,100},
        yticklabels={0,25,50,75,100},
        legend pos=north east,
        width=1.2\linewidth,
        legend style={font=\tiny, nodes={scale=0.7, transform shape}, at={(1.00,1.00)},anchor=north east},
        grid style=dashed
    ]

    %

    \addplot[color=color_orange, line width=1.2pt, mark=*] table[x index=0,y index=1,col sep=comma] {src/data/LowerConvexHull_FPUEnergy/Kmeans-CIP.csv};
    \addplot[color=color_brown, line width=1.2pt, mark=triangle*] table[x index=0,y index=1,col sep=comma] {src/data/LowerConvexHull_FPUEnergy/Kmeans-WP.csv};

    \legend{WP, CIP}


    \end{axis}

\end{tikzpicture}

\vspace{-0.1em}
         \label{fig:fpu_energy_Kmeans}
     \end{subfigure}
     \begin{subfigure}{.24\linewidth}
       \tikzset{mark options={mark size=1.4 pt}}
%
\pgfplotsset {every mark/.append style={line width=0.4pt}}
\pgfplotsset{every axis/.append style={font=\scriptsize}}

\newcommand{\axisheight}{0.54408}


\begin{tikzpicture}

    \begin{axis}[
        title={\small {Particlefilter}},
        title style={yshift= -0.5em},
        xlabel={Normalized Error (\%)},
        ylabel={NEC(\%)},
        xlabel style={yshift= 0.6em},
        y label style={at={(axis description cs:0.2,.5)}},
        xmin=0, xmax=20,
        ymin=0, ymax=100,
        xtick={0,5,10,15,20},
        xticklabels={0,5,10,15,20},
        ytick={0,25,50,75,100},
        yticklabels={0,25,50,75,100},
        legend pos=north east,
        width=1.2\linewidth,
        legend style={font=\tiny, nodes={scale=0.7, transform shape}, at={(1.00,1.00)},anchor=north east},
        grid style=dashed
    ]

    %

    \addplot[color=color_orange, line width=1.2pt, mark=*] table[x index=0,y index=1,col sep=comma] {src/data/LowerConvexHull_FPUEnergy/Particlefilter-CIP.csv};
    \addplot[color=color_brown, line width=1.2pt, mark=triangle*] table[x index=0,y index=1,col sep=comma] {src/data/LowerConvexHull_FPUEnergy/Particlefilter-WP.csv};

    \legend{WP, CIP}


    \end{axis}

\end{tikzpicture}

\vspace{-0.1em}
         \label{fig:fpu_energy_Particlefilter}
     \end{subfigure}
      \begin{subfigure}{.24\linewidth}
        \tikzset{mark options={mark size=1.4 pt}}
%
\pgfplotsset {every mark/.append style={line width=0.4pt}}
\pgfplotsset{every axis/.append style={font=\scriptsize}}

\newcommand{\axisheight}{0.54408}


\begin{tikzpicture}

    \begin{axis}[
        title={\small {Radar}},
        title style={yshift= -0.5em},
        xlabel={Normalized Error (\%)},
        ylabel={NEC(\%)},
        xlabel style={yshift= 0.6em},
        y label style={at={(axis description cs:0.2,.5)}},
        xmin=0, xmax=20,
        ymin=0, ymax=100,
        xtick={0,5,10,15,20},
        xticklabels={0,5,10,15,20},
        ytick={0,25,50,75,100},
        yticklabels={0,25,50,75,100},
        legend pos=north east,
        width=1.2\linewidth,
        legend style={font=\tiny, nodes={scale=0.7, transform shape}, at={(1.00,1.00)},anchor=north east},
        grid style=dashed
    ]

    %

    \addplot[color=color_orange, line width=1.2pt, mark=*] table[x index=0,y index=1,col sep=comma] {src/data/LowerConvexHull_FPUEnergy/Radar-CIP.csv};
    \addplot[color=color_brown, line width=1.2pt, mark=triangle*] table[x index=0,y index=1,col sep=comma] {src/data/LowerConvexHull_FPUEnergy/Radar-WP.csv};

    \legend{WP, CIP}


    \end{axis}

\end{tikzpicture}

\vspace{-0.1em}
         \label{fig:fpu_energy_Radar}
     \end{subfigure}
     \caption{Lower Convex Hulls of FPU energy and Error Rates for the WP and CIP. Values are normalized to the baseline.}
     \label{fig:fpu_energy}
     \vspace{-0.5em}
 \end{figure*}

Figure \ref{fig:fpu_energy} illustrates the lower convex hull of normalized FPU energy and the error rate (also referred to as accuracy loss). The error rate metric is the relative error of a configuration comparing against the highest quality configuration (baseline) where no approximation happens. The horizontal axis is the error rate while the Noramlized Energy Consumption (NEC) to the baseline is shown vertically (on the y-axis). The lower the curve is, the more efficient configuration is found which means higher energy efficiency. Since users generally do not care about extremely inaccurate outputs, only error rates less than 20\% is shown in the subfigures. The results show that f we assign multiple FPIs at the function level, NEAT will retrieve more energy efficient configurations that are not explorable if we use single FPI for the whole program.  This result further demonstrates NEAT's value in design space exploration.

With a minimal error in final output of the benchmark, NEAT reduces the FPU energy up to 60\%. For some applications such as \texttt{Blackscholes}, \texttt{Fluidanimate}, and \texttt{Particlefilter} the FPU energy savings are more considerable. These benchmarks have less than 10 FLOP intensive functions. Therefore first, CIP covers all the FLOPs in the program. Second, since the tradeoff space is relatively smaller, NSGA-II searches a larger portion of the tradeoff space in the same exploration time.




For \texttt{Fluidanimate} and \texttt{Ferret} benchmarks, there are only three and two configurations where the WP outperforms the CIP. The reason is that NEAT's genetic algorithm fails to explore those specific configurations as it is a heuristic algorithm. The same pattern can be seen for the \texttt{Radar} benchmark as well where the CIP does not dominate the whole-program approach.

The \texttt{Heartwall} benchmark has only two FLOP functions where they are very sensitive to the bit width adjustment and any modification leads to more than 20\% error. Consequently, NEAT is not able to decrease FPU energy to less than 71\% of the baseline with reasonable error rate. The opposite scenario happens for the \texttt{Particlefilter} application where the major FLOP functions do not impact the quality of output considerably, hence NEAT aggressively reduces the FPU energy without causing much error.

\begin{figure*}[tb]
  \begin{center}
\begin{tikzpicture}

    \pgfplotsset{
        height=4cm,
        width=0.95\linewidth,
        enlargelimits=false,
    }

    \pgfplotsset{every axis legend/.append style={
    draw=none,legend columns=6,at={(0.5,1.3)},anchor=north, column sep=1ex, font=\tiny},
    }


   \begin{axis}[
        axis x line* = bottom,
        major x tick style = transparent,
        xmin=0,
        xmax=10,
        xticklabel shift={0pt},
        x tick label style={ anchor=north}, 
        xtick={1,2,3,4,5,6,7,8,9},
        xticklabels={{\scriptsize $\mathsf{Blackscholes}$},
        {\scriptsize $\mathsf{Bodytrack}$},
        {\scriptsize $\mathsf{Fluidanimate}$},
        {\scriptsize $\mathsf{Ferret}$},
        {\scriptsize $\mathsf{Heartwall}$},
        {\scriptsize $\mathsf{Kmeans}$},
        {\scriptsize $\mathsf{Particlefilter}$},
        {\scriptsize $\mathsf{Radar}$},
        {\scriptsize $\mathsf{HarmonicMean}$}},
        ybar=\pgflinewidth,
        bar width=5,
        ymajorgrids,
        grid style={dashed},
        ylabel={FPU Energy Savings(\%)},
        y label style={font=\scriptsize, at={(axis description cs:0.01,.5)}},
        ymin=0,
        ymax=100,
        ytick={0,25,50,75,100},
        yticklabels={0,25,50,75,100},
      ]
        \addplot[color=color_orange,fill=color_orange] table[x index=0,y index=2, col sep=tab] {DatafpuEnergy.txt};
        \addplot[color=color_pink,fill=color_pink] table[x index=0,y index=3, col sep=tab] {DatafpuEnergy.txt};
        \addplot[color=color_purple,fill=color_purple] table[x index=0,y index=4, col sep=tab] {DatafpuEnergy.txt};

        \addplot[color=color_orange,fill=color_orange, pattern=crosshatch dots, pattern color=color_orange] table[x index=0,y index=5, col sep=tab] {DatafpuEnergy.txt};
        \addplot[color=color_pink,fill=color_pink, pattern=crosshatch dots, pattern color=color_pink] table[x index=0,y index=6, col sep=tab] {DatafpuEnergy.txt};
        \addplot[color=color_purple,fill=color_purple, pattern=crosshatch dots, pattern color=color_purple] table[x index=0,y index=7, col sep=tab] {DatafpuEnergy.txt};

        \addlegendentry{{\footnotesize $WP-1\%$}}
        \addlegendentry{{\footnotesize $WP-5\%$}}
        \addlegendentry{{\footnotesize $WP-10\%$}}
        \addlegendentry{{\footnotesize $CIP-1\%$}}
        \addlegendentry{{\footnotesize $CIP-5\%$}}
        \addlegendentry{{\footnotesize $CIP-10\%$}}





    \end{axis}


\end{tikzpicture}
    \caption{FPU Energy Savings at Different Error Rates, normalized to the baseline. Higher the bars, the more energy efficient is.}
    \label{fig:fpuEnergyErrorThresholds}
  \end{center}
\end{figure*}
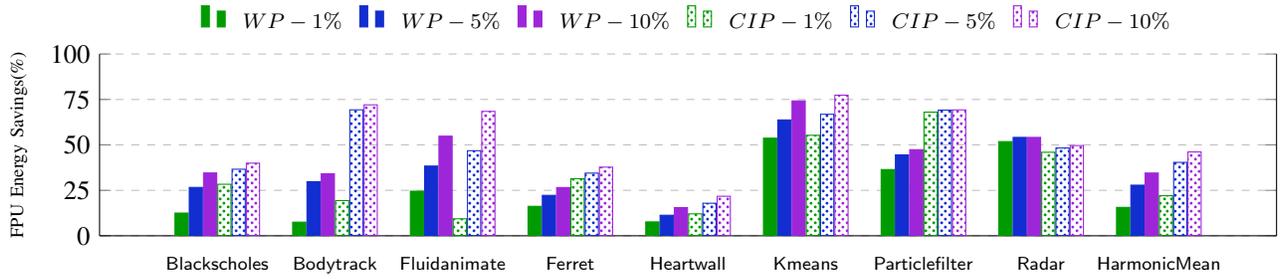

For a more detailed comparison, we re-illustrate a quantized representation of the previous plot. Figure \ref{fig:fpuEnergyErrorThresholds} displays how the FPU energy savings enhance as the tolerated error threshold increases. Higher bars indicate more energy savings. By harmonic mean, applying the CIP versus WP approach results in 7\%, 12\%, and 13\% more energy savings at 1\%,5\%, and 10\% error rate, respectively.

The steeper slope in the lower convex hull curves in subplots of Figure \ref{fig:fpu_energy} translates into higher bars in Figure \ref{fig:fpuEnergyErrorThresholds} as the error threshold increases. The \texttt{Blackscholes} and \texttt{Particlefilter} benchmarks demonstrate such behavior. On the contrary, by increasing the error threshold in \texttt{Particlefilter} and \texttt{Radar} applications, the FPU energy savings do not inflate similarly.

From these graphs, we draw two conclusions. First, specifying the FPIs placement at a finer granularity results in more efficient FPI to function mappings. In other words, per-function rules use less energy with the same error comparing to use a single FPI for the whole application. This type of insight is really only achievable with the an automated system like NEAT.  Second, if higher error rates are allowed, NEAT achieves higher efficiency of FPU energy.  Thus, NEAT can navigate the whole tradeoffs space and give users a range of options depending on tolerable error rate.

\subsection{Memory Instructions}
\label{ss:mem_insns}

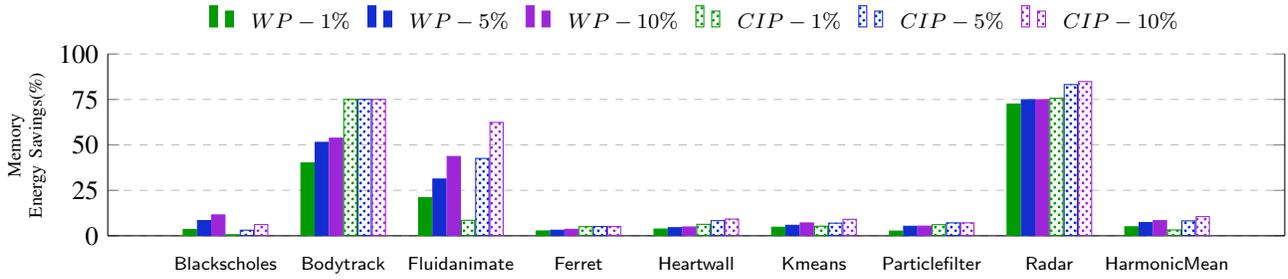
\begin{figure*}[tb]
  \begin{center}
\begin{tikzpicture}

    \pgfplotsset{
        height=4cm,
        width=0.95\linewidth,
        enlargelimits=false,
    }

    \pgfplotsset{every axis legend/.append style={
    draw=none,legend columns=6,at={(0.5,1.3)},anchor=north, column sep=1ex, font=\tiny},
    }


   \begin{axis}[
        axis x line* = bottom,
        major x tick style = transparent,
        xmin=0,
        xmax=10,
        xticklabel shift={0pt},
        x tick label style={anchor=north},
        xtick={1,2,3,4,5,6,7,8,9},
        xticklabels={{\scriptsize $\mathsf{Blackscholes}$},
        {\scriptsize $\mathsf{Bodytrack}$},
        {\scriptsize $\mathsf{Fluidanimate}$},
        {\scriptsize $\mathsf{Ferret}$},
        {\scriptsize $\mathsf{Heartwall}$},
        {\scriptsize $\mathsf{Kmeans}$},
        {\scriptsize $\mathsf{Particlefilter}$},
        {\scriptsize $\mathsf{Radar}$},
        {\scriptsize $\mathsf{HarmonicMean}$}},
        ybar=\pgflinewidth,
        bar width=5,
        ymajorgrids,
        grid style={dashed},
        ylabel={Memory \\ Energy Savings(\%)},
        y label style={align=center, font=\scriptsize, at={(axis description cs:0.01,.5)}},
        ymin=0,
        ymax=100,
        ytick={0,25,50,75,100},
        yticklabels={0,25,50,75,100},
      ]
        \addplot[color=color_orange,fill=color_orange] table[x index=0,y index=2, col sep=tab] {DatafpMemoryEnergy.txt};
        \addplot[color=color_pink,fill=color_pink] table[x index=0,y index=3, col sep=tab] {DatafpMemoryEnergy.txt};
        \addplot[color=color_purple,fill=color_purple] table[x index=0,y index=4, col sep=tab] {DatafpMemoryEnergy.txt};

        \addplot[color=color_orange,fill=color_orange, pattern=crosshatch dots, pattern color=color_orange] table[x index=0,y index=5, col sep=tab] {DatafpMemoryEnergy.txt};
        \addplot[color=color_pink,fill=color_pink, pattern=crosshatch dots, pattern color=color_pink] table[x index=0,y index=6, col sep=tab] {DatafpMemoryEnergy.txt};
        \addplot[color=color_purple,fill=color_purple, pattern=crosshatch dots, pattern color=color_purple] table[x index=0,y index=7, col sep=tab] {DatafpMemoryEnergy.txt};

        \addlegendentry{{\footnotesize $WP-1\%$}}
        \addlegendentry{{\footnotesize $WP-5\%$}}
        \addlegendentry{{\footnotesize $WP-10\%$}}
        \addlegendentry{{\footnotesize $CIP-1\%$}}
        \addlegendentry{{\footnotesize $CIP-5\%$}}
        \addlegendentry{{\footnotesize $CIP-10\%$}}





    \end{axis}


\end{tikzpicture}
    \caption{Memory Transfer Energy Savings at Different Error Rates, normalized to the baseline.}
    \label{fig:fpMemoryEnergyErrorThresholds}
  \end{center}
\end{figure*}

Main memory (DRAM) consumes as much as half of the total system power in a computer today, due to the increasing demand for memory capacity and bandwidth \cite{Malladi12}. Hence, reducing the memory traffic directly derives into substantial energy savings. NEAT estimates the memory energy with accounting only accesses to/from an off-chip memory by keeping track of memory operations such as MOVSS and MOVSD. Figure \ref{fig:fpMemoryEnergyErrorThresholds} depicts memory accesses energy for a range of error rates for both whole-program (WP) and per-function (CIP) approaches respectively across the benchmarks. Same as before, higher bars indicate higher energy efficiency. Values are normalized to non-approximated version of the application, that acts as a baseline. On harmonic mean, increasing the error rate from 1\% to 10\% results in 3.2-10.5\% less energy consumption.

If the FLOP functions are memory intensive, reducing the precision bits results in lower memory bandwidth, and consequently more energy savings. That is the reason why benchmarks such as \texttt{Bodytrack}, \texttt{Fluidanimate}, and \texttt{Radar} reduces the memory energy by more than 60\%. In rest of the benchmarks, the FLOP functions were solely compute intensive.

To put the experiments above to a conclusion, we illustrate the WP rule as a sample for prior work \cite{wu2018} which tries to find a single most optimal approximation for the whole application. The per-function rules of NEAT show off the ability of the replacement rules to allow programmers to explore a richer set of tradeoffs without having to come up with whole new implementations of existing program functionality.

\subsection{Flexible Precision Level}
\label{ss:flex_precision_level}

In previous sections, we observed some benchmarks have a mixture of both $float$ and $double$ FLOPs. To choose the right optimization target, we compare the energy and accuracy of selected benchmarks in both single and double optimization targets. The FPI to function mapping is CIP in this experiment.

\begin{figure}[tb]
  \begin{center}
\begin{tikzpicture}

    \pgfplotsset{
        height=4cm,
        width=0.95\columnwidth,
        enlargelimits=false,
    }

    \pgfplotsset{every axis legend/.append style={
    draw=none,legend columns=3,at={(0.5,1.5)},anchor=north, column sep=1ex, font=\tiny},
    }


   \begin{axis}[
        axis x line* = bottom,
        major x tick style = transparent,
        xmin=0,
        xmax=4,
        xticklabel shift={0pt},
        x tick label style={ anchor=north},
        xtick={1,2,3},
        xticklabels={{\scriptsize $\mathsf{Canneal}$},
        {\scriptsize $\mathsf{Ferret}$},
        {\scriptsize $\mathsf{Particlefilter}$}},
        ybar=\pgflinewidth,
        bar width=5.5,
        ymajorgrids,
        grid style={dashed},
        ylabel={FPU \\ Energy Savings(\%)},
        y label style={align=center, font=\scriptsize, at={(axis description cs:0.01,.5)}},
        ymin=0,
        ymax=100,
        ytick={0,25,50,75,100},
        yticklabels={0,25,50,75,100},
      ]
        \addplot[color=color_orange,fill=color_orange] table[x index=0,y index=2, col sep=tab] {DatafpuEnergyDualPrecisionOpt.txt};
        \addplot[color=color_pink,fill=color_pink] table[x index=0,y index=3, col sep=tab] {DatafpuEnergyDualPrecisionOpt.txt};
        \addplot[color=color_purple,fill=color_purple] table[x index=0,y index=4, col sep=tab] {DatafpuEnergyDualPrecisionOpt.txt};

        \addplot[color=color_orange,fill=color_orange, pattern=north east lines, pattern color=color_orange] table[x index=0,y index=5, col sep=tab] {DatafpuEnergyDualPrecisionOpt.txt};
        \addplot[color=color_pink,fill=color_pink, pattern=north east lines, pattern color=color_pink] table[x index=0,y index=6, col sep=tab] {DatafpuEnergyDualPrecisionOpt.txt};
        \addplot[color=color_purple,fill=color_purple, pattern=north east lines, pattern color=color_purple] table[x index=0,y index=7, col sep=tab] {DatafpuEnergyDualPrecisionOpt.txt};

        \addlegendentry{{\footnotesize $float-1\%$}}
        \addlegendentry{{\footnotesize $float-5\%$}}
        \addlegendentry{{\footnotesize $float-10\%$}}
        \addlegendentry{{\footnotesize $double-1\%$}}
        \addlegendentry{{\footnotesize $double-5\%$}}
        \addlegendentry{{\footnotesize $double-10\%$}}





    \end{axis}


\end{tikzpicture}
    \caption{FPU Energy Savings with Different Optimization Targets for NEAT.}
    \label{fig:fpuEnergyDualPrecisionOpt}
  \end{center}
\end{figure}
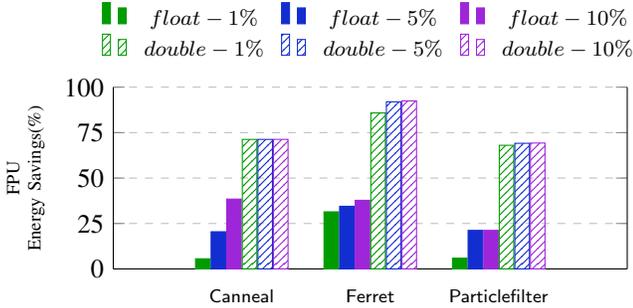


Figure \ref{fig:fpuEnergyDualPrecisionOpt} shows the normalized energy savings for both single and double optimization targets. As expected, if we choose the optimization target to be the same as the FP type which has larger ratio in FLOP distribution, higher energy savings would be achieved. This observation can be easily justified by the looking back at Section \ref{ss:fp_PD}. Both \texttt{Canneal} and \texttt{Particlefilter} contain more 64-bit than 32-bit FLOPs. Thus, double precision as NEAT directive is the right choice to achieve substantially higher energy efficiency.


\texttt{Ferret} requires special attention as it is not obvious how to choose the optimization target based on FLOP distribution ratio since it has almost equal amount of $float$ and $double$ FLOPs. At the 10\% error rate, NEAT saves up to 92\% of FPU energy corresponding to $double$ instructions while only 38\% savings is available if we consider only $float$ instructions. There  are  two reasons  for  the  discrepancy. One is that generally $double$ FLOPs yield more precise output, but they use more precision bits in return. Thus, NEAT has more freedom to cut down unnecessary floating point bits while not losing much accuracy because the $double$ baseline is already more accurate than the $float$ one. Second, the $double$ functions in \texttt{Ferret} are not accuracy sensitive, meaning that enforcing approximation on these functions would not excessively change the quality of the output.
This is a great example of how NEAT determines the most efficient configurations for any benchmark regardless of how their floating point precision is specified in the source (or binary).

\subsection{Function Call Stack}
\label{ss:func_call_stack}

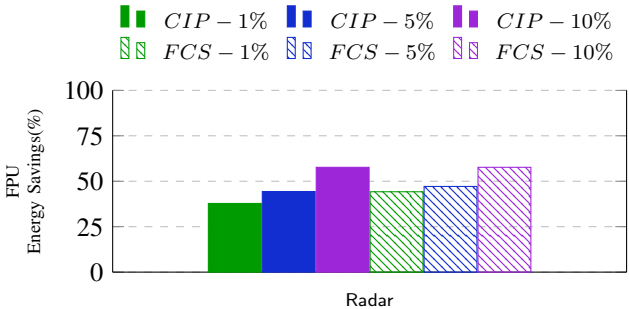
\begin{figure}[tb]
  \begin{center}
\begin{tikzpicture}

    \pgfplotsset{
        height=4cm,
        width=0.95\columnwidth,
        enlargelimits=false,
    }

    \pgfplotsset{every axis legend/.append style={
    draw=none,legend columns=3,at={(0.5,1.5)},anchor=north, column sep=1ex, font=\tiny},
    }


   \begin{axis}[
        axis x line* = bottom,
        major x tick style = transparent,
        xmin=0,
        xmax=2,
        xticklabel shift={0pt},
        x tick label style={anchor=north},
        xtick={1},
        xticklabels={{\scriptsize $\mathsf{Radar}$}},
        ybar=\pgflinewidth,
        bar width=20,
        enlarge x limits=1.0,
        ymajorgrids,
        grid style={dashed},
        ylabel={FPU \\ Energy Savings(\%)},
        y label style={align=center, font=\scriptsize, at={(axis description cs:0.01,.5)}},        ymin=0,
        ymax=100,
        ytick={0,25,50,75,100},
        yticklabels={0,25,50,75,100},
      ]
        \addplot[color=color_orange,fill=color_orange] table[x index=0,y index=2, col sep=tab] {DatafunctionCallStack.txt};
        \addplot[color=color_pink,fill=color_pink] table[x index=0,y index=3, col sep=tab] {DatafunctionCallStack.txt};
        \addplot[color=color_purple,fill=color_purple] table[x index=0,y index=4, col sep=tab] {DatafunctionCallStack.txt};

        \addplot[color=color_orange,fill=color_orange, pattern=north west lines, pattern color=color_orange] table[x index=0,y index=5, col sep=tab] {DatafunctionCallStack.txt};
        \addplot[color=color_pink,fill=color_pink, pattern=north west lines, pattern color=color_pink] table[x index=0,y index=6, col sep=tab] {DatafunctionCallStack.txt};
        \addplot[color=color_purple,fill=color_purple, pattern=north west lines, pattern color=color_purple] table[x index=0,y index=7, col sep=tab] {DatafunctionCallStack.txt};

        \addlegendentry{{\footnotesize $CIP-1\%$}}
        \addlegendentry{{\footnotesize $CIP-5\%$}}
        \addlegendentry{{\footnotesize $CIP-10\%$}}
        \addlegendentry{{\footnotesize $FCS-1\%$}}
        \addlegendentry{{\footnotesize $FCS-5\%$}}
        \addlegendentry{{\footnotesize $FCS-10\%$}}





    \end{axis}


\end{tikzpicture}
    \caption{Comparison of CIP and FCS for the FPU Energy Savings in Radar.}
    \label{fig:funcCallStak}
  \end{center}
\end{figure}

As we mentioned in section \ref{exec_flops}, if we map an FPI to a function, depending on the caller, the quality of output could change. While on most benchmarks, CIP and FCS approaches produce the same result, on the \texttt{Radar} they differ. Hence, we examine the impact of the caller of the FFT function on the energy and accuracy of the benchmark. Figure \ref{fig:funcCallStak} illustrates the FPU energy savings normalized to a baseline for CIP and FCS placement rules. FCS was able to explore a handful of more optimal configurations, resulting in 7\% more energy savings at 1\% accuracy loss comparing to CIP without extra runtime overhead. At 5\% and 10\% error rate, the additional energy savings are 4\% and 2\% respectively.

\subsection{Sensitivity to Input Changes}
\label{ss:coeff_cov}


Since we employ a heuristic exploration technique, we ensure that NEAT produces statistically sound results by evaluating each application with multiple inputs divided into training and test sets. We take the median of normalized accuracy loss and FPU energy for each set of inputs, compute a linear least squares fit of training data to test data, and compute the correlation coefficient of each
fit. Higher correlation coefficients imply less input sensitivity; i.e. the behavior of configurations found during training data is a good predictor of test behavior.

\begin{table}[tb]
  \centering
  \caption{Correlation Coefficients for Error Rates and FPU energy.}
\footnotesize
\begin{tabularx}{0.5\textwidth}{|C{1}|C{1}|C{1}|}
\hline
 \textbf{Benchmark } & \textbf{Error Rates} & \textbf{FPU Energy}\\
\hline
Blackscholes & 0.999 & 0.999\\
\hline
Bodytrack & 0.958 & 0.989\\
\hline
Fluidanimate & 0.995 & 1.0\\
\hline
Ferret & 0.973 & 1.0\\
\hline
Heartwall & 0.999 & 1.0\\
\hline
Kmeans & 0.932 & 1.0\\
\hline
Particlefilter  & 0.991 & 1.0\\
\hline
Radar  & 0.992 & 1.0\\
\hline
\end{tabularx}
\label{tbl:corr_coeff}
\end{table}

Table \ref{tbl:corr_coeff} show the correlation coefficient (R-values) for accuracy loss and FPU energy for each benchmark. Due to heuristic nature of exploration technique, it might be possible to select configurations that perform differently on unseen data. For instance, \texttt{Kmeans} clearly stresses the difference between training and test inputs. Although, all benchmarks have uniformly high R-values on accuracy loss and FPU energy---at least $0.93$.  This demonstrates that NEAT's search techniques are robust and the accuracy and energy results they predict on training inputs hold up well for test inputs.  The robustness of the energy results is, perhaps, not surprising as those should be highly predictable (simpler FLOP implementations should predictably lower energy).  The robustness of the accuracy results is perhaps more surprising as it not intuitively obvious that floating point implementations that work well for one set of inputs would also work for another set.  

\subsection{Neural Network Integration}
\label{ss:nn_integration_eval}


The energy and resource constraints in neural networks creates an intriguing challenge. More recently, a growing body of literature have tried to sacrifice the precision of training and inference for the lower runtime and energy consumption\cite{das2018}. NEAT can be used to identify the FLOP intensive sections of the network and then provide the minimum precision required for the computation without considerable model accuracy reductions. This tradeoff (small accuracy loss for large energy savings) is well known, and we perform this study not to claim a new result here, but to demonstrate that NEAT's automated approach can produce the same types of savings for this problem that have been produced by human domain experts.  We also believe that using NEAT's programmable replacement rules to create DNNs with differing precision throughout the network is a new contribution that would (due to the size of the search space) be quite difficult even for human experts.

We use a hand-written digit classification with the MNIST dataset which includes 60K images and 10K labels. For the CNN, we consider the LeNet-5 model with the architecture summary listed in Table \ref{tbl:lenet5_arch}. The LeNet-5 architecture consists of two sets of convolutional and average pooling layers, followed by a flattening convolutional layer, then two fully-connected layers and finally a softmax classifier\cite{lecun1999}.

\begin{table*}[ht]
  \centering
  \caption{LeNet-5 Architecture Summary.}
\footnotesize
\begin{tabularx}{0.95\textwidth}{C{0.9}|C{1.3}|C{0.9}|C{0.9}|C{1}|C{1}}
 \multicolumn{2}{c|}{\textbf{Layer}} & \textbf{Feature Map} & \textbf{Size} & \textbf{Kernel Size} & \textbf{Activation}\\
\hline
Input & Image & 1 & 32x32 & - & -\\
1 & Convolutional(1) & 6 & 28x28 & 5x5 & tanh\\
2 & Average Pooling(1) & 6 & 14x14 & 2x2 & tanh\\
3 & Convolutional(2) & 16 & 10x10 & 5x5 & tanh\\
4 & Average Pooling(2) & 16 & 5x5 & 2x2 & tanh\\
5 & Convolutional(3) & 120 & 1x1 & 5x5 & tanh\\
6 & Fully Connected & - & 84 & - & tanh\\
Output & Fully Connected & - & 10 & - & softmax\\
\end{tabularx}
\label{tbl:lenet5_arch}
\end{table*}

\newcolumntype{E}[1]{>{\hsize=#1\hsize\centering\tiny\arraybackslash}X}%
\newcolumntype{M}[1]{>{\centering\arraybackslash}m{#1}}

\begin{table*}[h!]
  \centering
  \caption{Mantissa Bits For Single Precision FP Recommended by NEAT for Each Layer at Different Error Rates.}
\footnotesize
\begin{tabularx}{0.95\textwidth}{|C{1}|C{1}|C{1}|C{1}|C{1}|C{1}|C{1}|C{1}|C{1}|}
\hline
 \textbf{Layers / Error Rates } & \textbf{Conv 1} & \textbf{Avg Pool 1} & \textbf{Conv 2} & \textbf{Avg Pool 2} & \textbf{Conv 3} & \textbf{FC} & \textbf{Tanh} & \textbf{Internal Func.}   \\
\hline
1 \% & 10 & 23 & 14 & 4 & 19 & 4 & 20 & 17\\
\hline
5 \% & 10 & 5 & 5 & 16 & 13 & 4 & 18 & 15\\
\hline
10 \% & 6 & 16 & 12 & 9 & 13 & 1 & 17 & 11\\
\hline
\end{tabularx}
\label{tbl:pred_precison_bits}
\end{table*}

\begin{figure}[t]
    \centering
    \begin{subfigure}{0.95\linewidth}
        \includegraphics[width=0.95\textwidth]{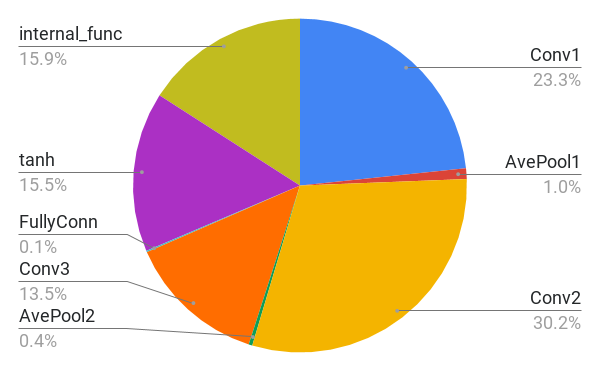}
    \end{subfigure}
    \caption{32-bit FLOP breakdown per layer in digit recognition CNN.}
    \label{fig:mnist_flopBreakdown}
    \vspace{-1.5em}
\end{figure}

 Figure \ref{fig:mnist_flopBreakdown} shows the FLOPs breakdown for CNN training with minibatch size of 4, learning rate of 1, and 30 epochs. We first measured how much of the operations are floating point to determine the applicability of NEAT. For the inference, more than 73\% of operations were FLOPs which makes NEAT absolutely beneficial to apply. Next, we analyze the FLOP distribution between the layers. We observe that more than 69\% of floating point computation happens in the convolutional layers where they extract interesting features in an image. Activation phases and internal compute functions are responsible for the majority of remainder. Finally, we show that the number of FLOPs decreases as we approach the latter layers of the CNN since the size of transferred data between layers reduces as well.

To apply the FPI to function placement rules for a CNN, there are two options. First, apply one FPI per layer category (we refer to as PLC) meaning that all convolutional layers use the same precision level. The second approach is to apply a different FPI Per Layer Instance (PLI) where in this case the first and third layers might use distinct precision levels, however, they are both convolutional layers.

Picking the right FPI placement policy is not trivial for the CNNs. Unlike the WP versus CIP rules where one has significantly larger tradeoff space, the PLC and PLI tradeoff spaces are both large enough that heuristic exploration is required. Thus, any of these rules could outperform the other with the same exploration time. For the PLC, NEAT explores a larger portion of the tradeoff space, leading to locating efficient configurations more quickly. On the other hand, PLI examines FPI mappings at a finer granularity, hence it has a higher chance of discovering more optimal configurations.

\begin{figure}
  \begin{subfigure}[b]{0.53\columnwidth}
    \fcolorbox{white}{white}{\hspace*{-27pt} \tikzset{mark options={mark size=1.4 pt}}
%
\pgfplotsset {every mark/.append style={line width=0.4pt}}
\pgfplotsset{every axis/.append style={font=\scriptsize}}

\newcommand{\axisheight}{0.54408}


\begin{tikzpicture}

    \begin{axis}[
        title={\small {CNN}},
        title style={yshift= -0.5em},
        xlabel={Normalized Error (\%)},
        ylabel={NEC(\%)},
        xlabel style={yshift= 0.6em},
        y label style={at={(axis description cs:0.2,.5)}},
        xmin=0, xmax=20,
        ymin=0, ymax=100,
        xtick={0,5,10,15,20},
        xticklabels={0,5,10,15,20},
        ytick={0,25,50,75,100},
        yticklabels={0,25,50,75,100},
        legend pos=north east,
        width=1.2\linewidth,
        legend style={font=\tiny, nodes={scale=0.7, transform shape} , at={(1.00,1.00)},anchor=north east},
        grid style=dashed
    ]

    %

    \addplot[color=color_orange, line width=1.2pt, mark=*] table[x index=0,y index=1,col sep=comma] {src/data/LowerConvexHull_FPUEnergy/TinyDNN-PLC.csv};
    \addplot[color=color_brown, line width=1.2pt, mark=triangle*] table[x index=0,y index=1,col sep=comma] {src/data/LowerConvexHull_FPUEnergy/TinyDNN-PLI.csv};

    \addlegendentry{{\tiny $PLC$}}
    \addlegendentry{{\tiny $PLI$}}


    \end{axis}

\end{tikzpicture}

\vspace{-0.1em}}
    \caption{}
    \label{fig:tinyDNNLCH}
  \end{subfigure}
  \begin{subfigure}[b]{0.45\columnwidth}
    \fcolorbox{white}{white}{\hspace*{-15pt} 
\begin{tikzpicture}

    \pgfplotsset{
        height=4cm,
        width=\columnwidth,
        enlargelimits=false,
    }



   \begin{axis}[
        axis x line* = bottom,
        width = 1.2\linewidth,
        major x tick style = transparent,
        xmin=0,
        xmax=2,
        xticklabel shift={0pt},
        xtick={1},
        xticklabels={{\scriptsize $\mathsf{CNN}$}},
        ybar=\pgflinewidth,
        bar width=8,
        ymajorgrids,
        grid style={dashed},
        ylabel={FPU Energy Savings(\%)},
        y label style={align=center, font=\scriptsize, at={(axis description cs:0.20,.5)}},
        y tick label style={font=\tiny},
        ymin=0,
        ymax=100,
        ytick={0,25,50,75,100},
        yticklabels={0,25,50,75,100},
        legend style = {draw=none,legend columns=2,at={(1.05,1)},anchor=south east, column sep=1ex, nodes={scale=0.65, transform shape} , font=\tiny}
      ]
        \addplot[color=color_orange,fill=color_orange] table[x index=0,y index=2, col sep=tab] {DatafunctionCallStack.txt};
        \addplot[color=color_pink,fill=color_pink] table[x index=0,y index=3, col sep=tab] {DatafunctionCallStack.txt};
        \addplot[color=color_purple,fill=color_purple] table[x index=0,y index=4, col sep=tab] {DatafunctionCallStack.txt};

        \addplot[color=color_orange,fill=color_orange, pattern=north west lines, pattern color=color_orange] table[x index=0,y index=5, col sep=tab] {DatafunctionCallStack.txt};
        \addplot[color=color_pink,fill=color_pink, pattern=north west lines, pattern color=color_pink] table[x index=0,y index=6, col sep=tab] {DatafunctionCallStack.txt};
        \addplot[color=color_purple,fill=color_purple, pattern=north west lines, pattern color=color_purple] table[x index=0,y index=7, col sep=tab] {DatafunctionCallStack.txt};

        \addlegendentry{{\tiny $PLC-1\%$}}
        \addlegendentry{{\tiny $PLC-5\%$}}
        \addlegendentry{{\tiny $PLC-10\%$}}
        \addlegendentry{{\tiny $PLI-1\%$}}
        \addlegendentry{{\tiny $PLI-5\%$}}
        \addlegendentry{{\tiny $PLI-10\%$}}





    \end{axis}


\end{tikzpicture}}
    \caption{}
    \label{fig:tinyDNNBarChart}
  \end{subfigure}
  \caption{Comparison of PLC and PLC replacements for the CNN. (a) Lower Convex Hull Curves of Energy and Error Rate. (b) Quantized Energy Savings at Different Error Rates.}
  \label{fig:tinyDNNMNIST}
\end{figure}
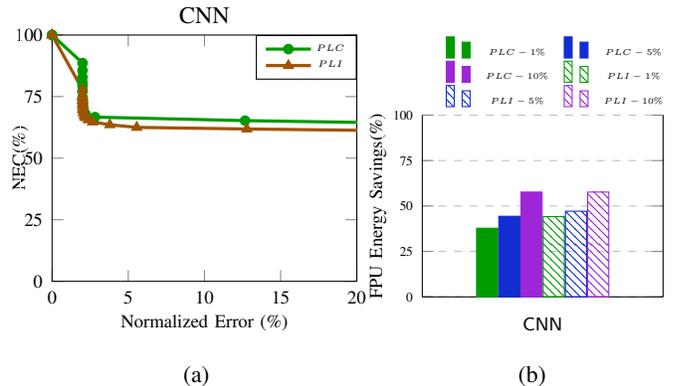


Figure \ref{fig:tinyDNNLCH} illustrates the lower convex hull of normalized FPU energy and accuracy for both approaches. The accuracy loss is the error difference to the baseline configuration without approximation. The baseline recognition accuracy in the inference stage is 99.04\% with a full accurate trained model. Each point in the tradeoff space demonstrates an FPI to layer (category or instance) mapping. Closer points to the origin indicate higher energy efficiency.

As can be seen, the lower convex hull of the PLI (finer granularity) outperforms the PLC curve for the error rates of less than 20\%. The quantized representation of FPU energy versus error rates tradeoff space is shown in Figure \ref{fig:tinyDNNBarChart} for both PLC and PLI placements. Similar to previous evaluation, finer granularity results in higher energy efficiency. With 1\%, 5\%, and 10\% accuracy loss, NEAT with PLI placements achieves 6\%, 4\%, and 3\% more energy savings compared to the default configuration.

NEAT's programmable placement rules allow developers to analyze various precision levels for different components of their neural network without requiring them to instrument the source code or re-design the architecture.

Since the FPIs are based on the bit truncation of mantissa, using the above analysis, NEAT finds the required precision bits for each layer in the LeNet-5 network under accuracy loss constraints. By default, each layer is implemented with single precision floating point numbers (24 mantissa bits) bits. Table \ref{tbl:pred_precison_bits} demonstrates the mantissa bits required for every layer in the network. These precisions could later be integrated with the MPFR library in C \cite{fousse2007} or mpmath library in Python \cite{Johansson2010}.

\section{Conclusion}
In this work, we proposed NEAT, a tool for automated precision tuning of floating point applications.  NEAT
provides mechanisms for programmers trying to explore the tradeoff space of combinations of approximate floating point implementations without extensive source code refactoring.
We evaluate NEAT on various benchmarks with whole-program and per-function placement rules. We found out at the finer granularity, up to 54\% and 74\% energy savings are available in FPU and memory transmissions respectively. We empirically show that NEAT performs robustly on unseen inputs as well. We also perform a case study on a digit recognition CNN programs to find optimal precision level requirements for each layer.

\paragraph*{Acknowledgments}
This research is supported by NSF(CCF-1439156, CNS-1526304, CCF-1823032, CNS-1764039). Additional support comes from the Proteus project under the DARPA BRASS program and a DOE Early Career award.





\bibliographystyle{IEEEtranS}
\bibliography{refs}

\end{document}